# A probabilistic model of the electron transport in films of PbSe nanocrystals arranged in a cubic lattice

Ilka Kriegel[1], Francesco Scotognella[2]*
[1]Department of Nanochemistry, Istituto Italiano di Tecnologia (IIT), via Morego, 30, 16163 Genova, Genova, Italy
[2]Dipartimento di Fisica, Politecnico di Milano, Piazza Leonardo da Vinci 32, 20133 Milano, Italy
* Corresponding author at: Dipartimento di Fisica, Politecnico di Milano, Piazza Leonardo da Vinci 32, 20133 Milano, Italy.
E-mail address: francesco.scotognella@polimi.it (F. Scotognella)

**Abstract**
The fabrication of nanocrystal (NC) films, starting from colloidal dispersion, is a very attractive topic in condensed matter physics community. NC films can be employed for transistors, light emitting diodes, laser, and solar cells. For this reason the understanding of the film conductivity is of major importance. In this paper we describe a probabilistic model that allow to predict the conductivity of the NC films, in this case of a cubic lattice of Lead Selenide NCs. The model is based on the hopping probability between NCs show a comparison with experimental data reported in literature.

**Introduction**
The interest on the electronic and optical properties of nanocrystal (NC) films is significantly increasing in the last years. NC films combine the easy processability of ink-based films with the outstanding physical properties of semiconductor nanocrystals[1]. The boundless chemical and physico-chemical effort to make NC films of different compounds ensuring charge mobility along them is testified by several works of many research groups[2–4].
However, charge carrier mobility among NCs is also very important in NC-polymer or NC-molecule hybrid films, which are employed in hybrid solar cells, as in Giansante et al.[5] and Mastria et al.[6]
The research group of Guyot-Sionnest presented an elegant description of a hopping model for the charge mobility in CdSe cubic superlattices[7]. Carbone, Carter, and Zimanyi describe a Monte Carlo simulation to study hole and electron mobility and electron conductivity in PbSe films[8]. Besides films based on semiconductor NCs, Ederth et al. presented a model for the optical and electronic properties of indium tin oxide nanoparticle films[9], where indium tin oxide nanoparticles show a plasmonic behaviour[10].
Here we describe a probabilistic model useful to predict the electron conductivity of a PbSe NC film. We studied the electron conductivity as a function of the size of the nanocrystals. The comparison between our model and the experimental results reported in literature show a good agreement. The beauty of the presented model is that the path of only one carrier is taken into account, while for other possibly present carriers a probability function has been implemented. This makes the approach attractive because of the low computational cost and can be implemented in scenarios in which charge carrier transfer, followed by transport, occurs is hybrid conducting NC films.

**Outline of the model**
We consider different cubic superlattices of PbSe nanocrystals, in which the parameter that changes is the size of the NCs (diameter) ranging from 3 to 8 nm. We take into account only the nearest neighbour hopping. To quantify the hopping rate between the site *1* and the site *2*

(being *1* and *2* two nanocrystals of the film), we refer to the work reported by Carbone, Carter, and Zimanyi[8] and references therein. In particular, the hopping rate *P* is

$$P_{1\to 2} = exp\left(\frac{-\Delta E + el\vec{F}\cdot\hat{i}_{12}}{k_B T}\right) \quad (1)$$

if the exponent of *P* is <0. If the exponent of *P* is ≥0, *P*=1. In the equation *e* is the electron charge, *l* is the hopping length (diameter of the PbSe NC, from 3 to 8 nm, and ligand length, 0.44 nm as in Ref. [8]), $k_B$ is the Boltzmann constant, T the temperature. In the model we used the same temperature of the experiments performed by Kang et al.[11], i.e. 200 K. The dot product is maximum when the hop direction is parallel to the electric field *F*.

*ΔE* refers to the difference in energy between the final level and the initial level, $E_2$-$E_1$. Following the calculations of An et al.[2] and the model of Carbone et al.[1], the energy of the initial site is

$$E_1 = E_{1,0} + (N_1 - 1)\left(\frac{X_C}{d}\right) \quad (2)$$

while the energy of the final site is

$$E_2 = E_{2,0} + N_2\left(\frac{X_C}{d}\right) \quad (3)$$

$E_{1,0}$ and $E_{2,0}$ are the lowest electron level for PbSe, as studied in Kang and Wise[13] and in Carbone et al.[8]. $X_C/d$ is the charging energy with $X_C$ the charging parameter ($X_C$ = 827.26 meV·nm as in Carbone et al.[8]) $N_1$ and $N_2$ are the number of electrons on the quantum dot site 1 and 2, respectively. Due to symmetry and spin degeneracy of the lowest electron level of PbSe, there can be eight electrons in a PbSe site[8,12]. In our model, based on the percolation path of a single electron, we quantified $N_1$ and $N_2$ in order to have a high probability to have $N_1$=1 and $N_2$=0, and a decreasing probability to have $N_1$>1 and $N_2$>0.

The mean value of the conductivity of the film can be written as

$$\langle\sigma\rangle = C\frac{l_{tot}}{A_{tot}}\langle P\rangle \quad (4)$$

being *C* a constant (in Siemens), $l_{tot}$ is the length of the NC film, $A_{tot}$ is the NC film cross section, *<P>* is

$$\langle P\rangle = \frac{1}{\langle N\rangle} \quad (5)$$

where *<N>* is the mean value of the number of attempted steps to reach the final electrode. For each electron that starts from its percolation path (from the starting electrode to the final electrode, see Figure 1), *N* is a natural positive number that increases whenever the hopping from a site to another site is successful or not (following Equation 1). To perform a statistical analysis of the percolation path, we do this simulation 500 times.

**Results and Discussion**

The simulation takes into account the hopping of an electron which travels from the starting electrode (corresponding to *x*=0) to the final electrode (corresponding to *x*=200). In Figure 1 we show a sketch of the "electron walk" along the cubic superlattice of a PbSe NC film (in the figure we truncated the plot to *x*=30). We underline that in this simulation we only consider the hopping between nearest neighbours.

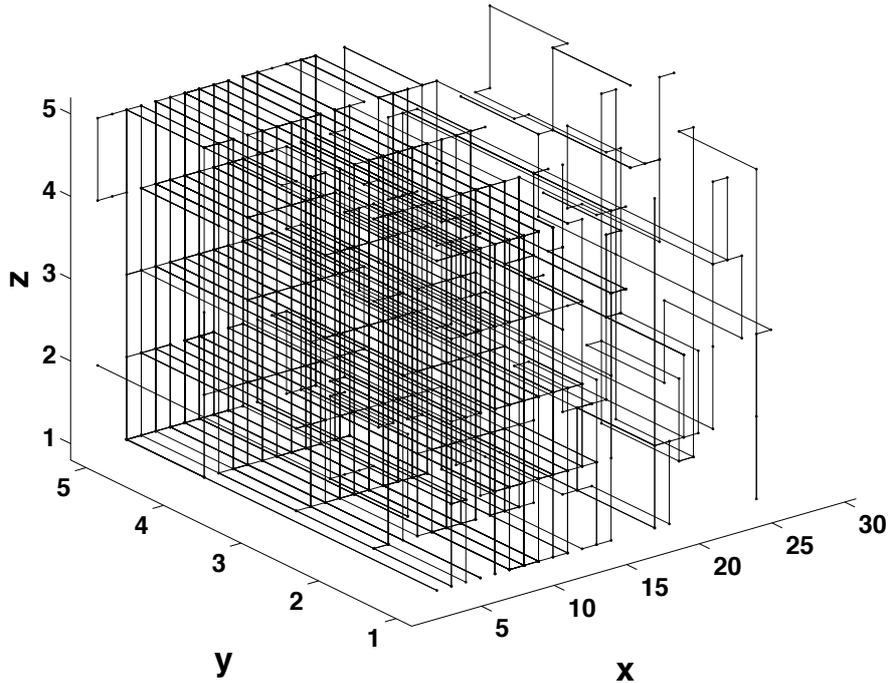

**Figure 1.** Example of a percolation path of an electron in a PbSe NC film.

In Figure 1 only the successful steps are depicted, underlining that the successful hopping probability follows the Equation 1. In Figure 2 we show the histogram (the calculation for a single electron has been reiterated 500 times) of the conductivity for a film of 8 nm large PbSe nanocrystals. In agreement with Equation 4, we found the parameter $C$ that relates the *<P>* with *<σ>*. In our work the parameter $C$ has been found to be equal to $1.5 \times 10^7$ S.

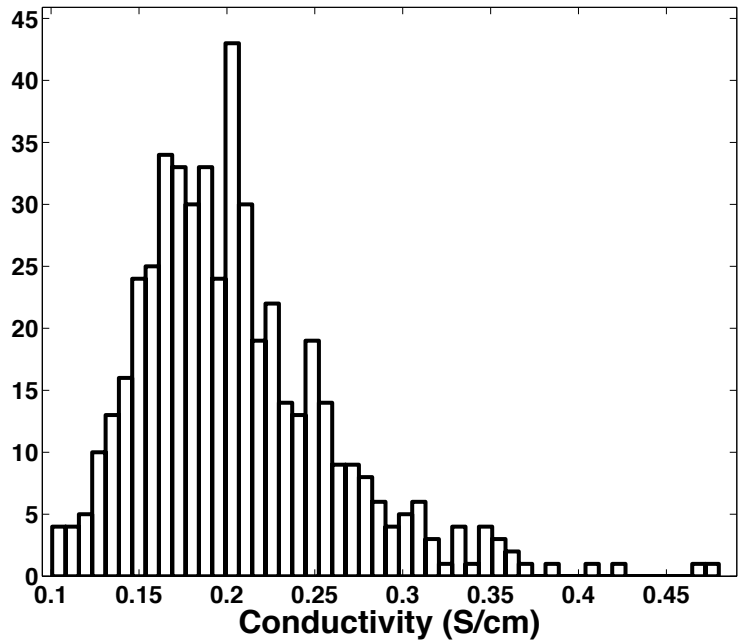

**Figure 2.** Histogram of the conductivity of a PbSe nanocrystal film in which the nanocrystals have diameter of 8 nm.

We have studied the conductivity as a function of the diameter of the NCs, from 3 to 8 nm. In Figure 3 we show the trend of the conductivity, with the standard deviation for each NC diameter. Each point corresponds to a distribution, as reported for 8 nm in Figure 2, and for all the diameters the distribution is skewed.

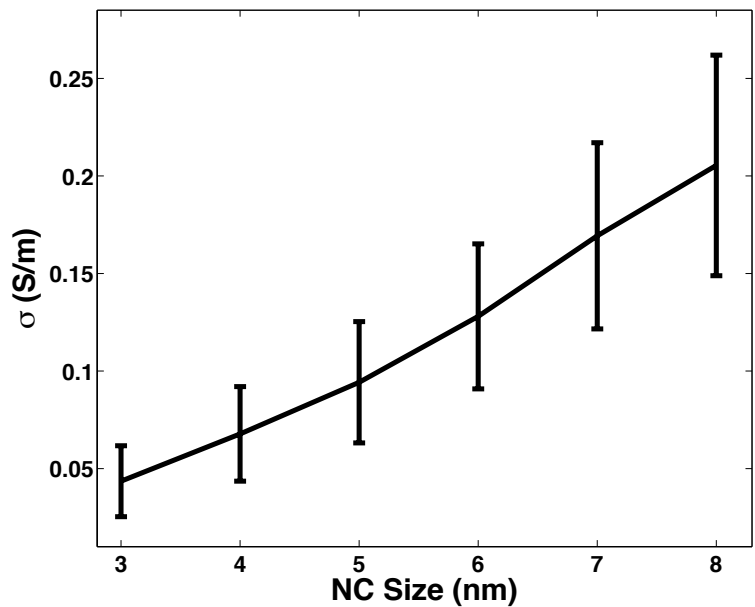

**Figure 3.** Conductivity of PbSe nanocrystal films as a function of the nanocrystal diameter

To find out the agreement between our simulations and the experimental results, we show in Figure 4 such comparison (the experimental data are taken from Kang et al.[3]). This comparison is dependent on the $C$ parameter. However, the trend of the simulation is in good agreement with the experimental results.

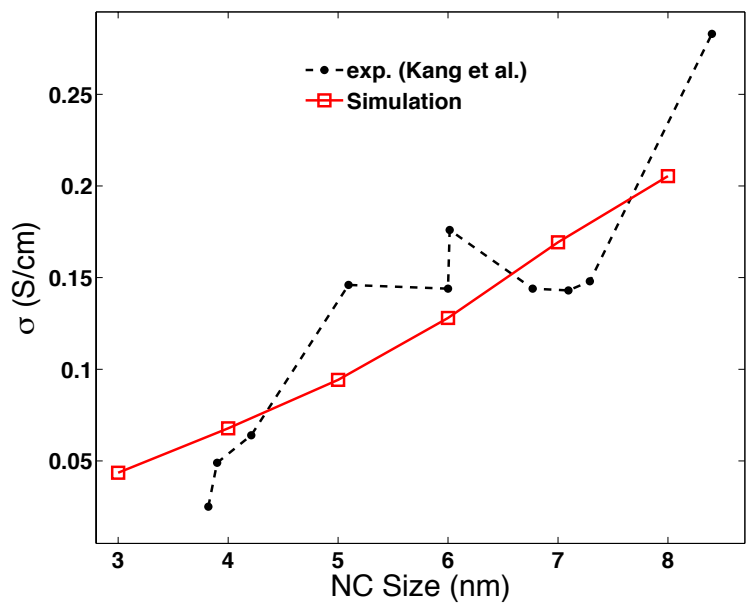

**Figure 4.** Conductivity of PbSe nanocrystal films as a function of the nanocrystal diameter: experimental values (black dots with dashed line, from Kang et al.[3]) and simulations (red squares with full line).

We would like to underline that the simulation is quite efficient in terms of computational costs (taking also into account that 500 simulations is well above the convergence) and the only parameter is *C*.

**Conclusions**

In this paper we describe a probabilistic model that allow predicting the conductivity of the NC films. In particular, we investigated the electron transport in a cubic lattice of Lead Selenide NCs. We show a good agreement with experimental data reported in literature.

The study is helpful for the understanding of the transport mechanisms in experimental NC films. This model can be improved to consider also amorphous NC superlattices. and might include the influence of the NC ligands on the transport[14]. Moreover, a further analysis could include the charge transport in more complicated systems, as for example NC-polymer heterojunctions[5].

**References**
(1) Talapin, D. V. *Science* **2005**, *310* (5745), 86–89.
(2) Talapin, D. V.; Lee, J.-S.; Kovalenko, M. V.; Shevchenko, E. V. *Chem. Rev.* **2010**, *110* (1), 389–458.
(3) Giansante, C.; Carbone, L.; Giannini, C.; Altamura, D.; Ameer, Z.; Maruccio, G.; Loiudice, A.; Belviso, M. R.; Cozzoli, P. D.; Rizzo, A.; Gigli, G. *Thin Solid Films* **2014**, *560*, 2–9.
(4) Baranov, A. V.; Ushakova, E. V.; Golubkov, V. V.; Litvin, A. P.; Parfenov, P. S.; Fedorov, A. V.; Berwick, K. *Langmuir* **2015**, *31* (1), 506–513.
(5) Giansante, C.; Mastria, R.; Lerario, G.; Moretti, L.; Kriegel, I.; Scotognella, F.; Lanzani, G.; Carallo, S.; Esposito, M.; Biasiucci, M.; Rizzo, A.; Gigli, G. *Adv. Funct. Mater.* **2015**, *25* (1), 111–119.
(6) Mastria, R.; Rizzo, A.; Giansante, C.; Ballarini, D.; Dominici, L.; Inganäs, O.; Gigli, G. *J. Phys. Chem. C* **2015**, *119* (27), 14972–14979.
(7) Yu, D.; Wang, C.; Wehrenberg, B. L.; Guyot-Sionnest, P. *Phys. Rev. Lett.* **2004**, *92* (21).
(8) Carbone, I.; Carter, S. A.; Zimanyi, G. T. *J. Appl. Phys.* **2013**, *114* (19), 193709.
(9) Ederth, J.; Heszler, P.; Hultåker, A.; Niklasson, G. .; Granqvist, C. . *Thin Solid Films* **2003**, *445* (2), 199–206.
(10) Scotognella, F.; Della Valle, G.; Srimath Kandada, A. R.; Zavelani-Rossi, M.; Longhi, S.; Lanzani, G.; Tassone, F. *Eur. Phys. J. B* **2013**, *86* (4).
(11) Kang, M. S.; Sahu, A.; Norris, D. J.; Frisbie, C. D. *Nano Lett.* **2011**, *11* (9), 3887–3892.
(12) An, J. M.; Franceschetti, A.; Zunger, A. *Phys. Rev. B* **2007**, *76* (4).
(13) Kang, I.; Wise, F. W. *J. Opt. Soc. Am. B* **1997**, *14* (7), 1632.
(14) Borriello, C.; Miscioscia, R.; Mansour, S. A.; Di Luccio, T.; Bruno, A.; Loffredo, F.; Villani, F.; Minarini, C. *Phys. Status Solidi A* **2015**, *212* (12), 2677–2685.